\documentclass[preprintnumbers,amsmath,amssymb,floatfix,11pt,prd,twocolumn,superscriptaddress,nofootinbib]{revtex4}
\usepackage{hyperref}
\usepackage{footnote}
\usepackage{amssymb}
\usepackage{amsmath}
\usepackage{graphicx}
\usepackage{dcolumn}
\usepackage{bm}
\usepackage{color}
\usepackage{xcolor}
\usepackage{float}
\usepackage{lipsum}
\allowdisplaybreaks
\usepackage{subfig}
\usepackage{comment}
\usepackage{multirow}
\usepackage[figurename=FIGURE,tablename=TABLE]{caption}
\usepackage{threeparttable}
\usepackage{hyperref}
\hypersetup{
    colorlinks=true, 
    linktoc=all,     
    linkcolor=magenta,
    citecolor=red
}

\newcommand{\ba}{\begin{eqnarray}}
\newcommand{\ea}{\end{eqnarray}}

\begin{document}

\title{Radiative-Corrected Higgs Inflation in Light of the Latest ACT Observations}

\author{Jureeporn Yuennan}
\email{jureeporn\_yue@nstru.ac.th}
\affiliation{Faculty of Science and Technology, Nakhon Si Thammarat Rajabhat University, Nakhon Si Thammarat, 80280, Thailand}

\author{Farruh Atamurotov}
\email{atamurotov@yahoo.com}
\affiliation{Urgench State University, Kh. Alimdjan str. 14, Urgench 220100, Uzbekistan}

\author{Phongpichit Channuie}
\email{phongpichit.ch@mail.wu.ac.th}
\affiliation{School of Science \& College of Graduate Studies, Walailak University, Nakhon Si Thammarat, 80160, Thailand}

\date{\today}

\begin{abstract}
Recent measurements from the Atacama Cosmology Telescope (ACT), particularly when combined with DESI baryon acoustic oscillation data, have reported a scalar spectral index $n_s$ slightly higher than that inferred by {\it Planck}~2018, suggesting a mild tension with the predictions of standard inflationary attractor models. In this work, we revisit the quantum-corrected Higgs inflation scenario within the framework of a non-minimally coupled scalar field theory. Starting from the one-loop effective action, we incorporate radiative corrections through the anomalous scaling parameter ${\bf A_I}$ and derive analytic expressions for the inflationary observables $n_s$ and $r$ in the Einstein frame. Our analysis demonstrates that quantum corrections naturally shift $n_s$ toward higher values while keeping the tensor-to-scalar ratio $r$ suppressed. For ${\cal N} = 60$, the model predicts $n_s \simeq 0.9743$ and $r \simeq 5.4\times10^{-3}$, in excellent agreement with the latest ACT+DESI (P-ACT-LB) data and fully consistent with the \textit{Planck}~2018 limit $r < 0.036$. The derived constraint $4.36\times10^{-10} < \lambda/\xi^{2} < 10.77\times10^{-10}$ confirms the robustness of the quantum-corrected Higgs framework and indicates that near-future CMB polarization experiments such as CORE, AliCPT, LiteBIRD, and CMB-S4 will be able to probe the predicted parameter space with high precision.
\end{abstract}


\maketitle

\newpage
\section{Introduction}

Recently, the Atacama Cosmology Telescope (ACT) data~\cite{ACT:2025fju,ACT:2025tim}, 
when analyzed in conjunction with the DESI survey~\cite{DESI:2024uvr,DESI:2024mwx}, 
have motivated the cosmology community to reassess the standard inflationary paradigm. 
The new ACT results indicate that the scalar spectral index of primordial curvature perturbations 
shows at least a $2\sigma$ discrepancy with the {\it Planck} 2018 findings~\cite{Planck:2018jri}, 
suggesting that refinements to the conventional inflationary framework may be required. 
Inflation remains a fundamental component of modern cosmology, providing elegant solutions 
to the flatness, horizon, and monopole problems of the Big Bang model. Moreover, 
it naturally accounts for the origin of primordial fluctuations that seeded the formation 
of cosmic structures and appear today as anisotropies in the cosmic microwave background (CMB)~\cite{Starobinsky:1980te,Sato:1981qmu,Guth:1980zm,Linde:1981mu,Albrecht:1982wi}. 
These perturbations are typically described by two observables: the scalar spectral index, $n_s$, 
characterizing the scale dependence of scalar modes, and the tensor-to-scalar ratio, $r$, 
which quantifies the relative amplitude of primordial gravitational waves.

In standard inflationary models, both $n_s$ and $r$ can be expressed in terms of 
the number of $e$-foldings, ${\cal N}$, between horizon exit and the end of inflation, 
enabling precise confrontation with observational data. A well-known and robust prediction 
among many models is the ``universal attractor'' relation $n_s = 1 - 2/{\cal N}$, 
which appears in a wide variety of scenarios, including $\alpha$-attractors~\cite{Kallosh:2013tua,Kallosh:2013hoa,Kallosh:2013maa,Kallosh:2013yoa,Kallosh:2014rga,Kallosh:2015lwa,Roest:2015qya,Linde:2016uec,Terada:2016nqg,Ueno:2016dim,Odintsov:2016vzz,Akrami:2017cir,Dimopoulos:2017zvq}, 
the Starobinsky $R^2$ model~\cite{Starobinsky:1980te}, 
and Higgs inflation with a large non-minimal coupling to gravity~\cite{Kaiser:1994vs,Bezrukov:2007ep,Bezrukov:2008ej}. The authors of Ref.\cite{Chang:2019ebx} discuss the one-loop corrections at finite temperature to the curvature perturbation generated during the Higgs inflation and demonstrate that thermal-loop effects give the Higgs inflation with a better fit to Planck CMB data. Comparable predictions also arise in models with composite inflaton fields~\cite{Karwan:2013iph,Channuie:2012bv,Bezrukov:2011mv,Channuie:2011rq}, 
as reviewed in~\cite{Channuie:2014ysa,Samart:2022pza}. 
For the benchmark value ${\cal N} = 60$, the universal relation predicts $n_s \approx 0.9667$, 
in excellent agreement with the {\it Planck} 2018 result $n_s = 0.9649 \pm 0.0042$~\cite{Planck:2018jri}.

However, the most recent ACT measurements~\cite{ACT:2025fju,ACT:2025tim} 
suggest a higher scalar spectral index. 
The combined ACT–{\it Planck} (P-ACT) analysis gives $n_s = 0.9709 \pm 0.0038$, 
and when the CMB lensing and BAO data from DESI are included (P-ACT-LB), 
the value further increases to $n_s = 0.9743 \pm 0.0034$. 
These new results place substantial tension on the universal attractor models, 
ruling them out at roughly the $2\sigma$ level and challenging even the Starobinsky $R^2$ model itself~\cite{ACT:2025fju}. More specifically, the authors of Ref.\cite{Gialamas:2025kef} first address Higgs-like inflation with radiative corrections to the inflationary potential in light of the ACT data. This outcome, both unexpected and significant, has inspired numerous theoretical studies 
aimed at reconciling inflationary predictions with the ACT data~\cite{Kallosh:2025rni,Gao:2025onc,Liu:2025qca,Yogesh:2025wak,Yi:2025dms,Peng:2025bws,Yin:2025rrs,Byrnes:2025kit,Wolf:2025ecy,Aoki:2025wld,Gao:2025viy,Zahoor:2025nuq,Ferreira:2025lrd,Mohammadi:2025gbu,Choudhury:2025vso,Odintsov:2025eiv,Odintsov:2025wai,Q:2025ycf,Zhu:2025twm,Kouniatalis:2025orn,Hai:2025wvs,Dioguardi:2025vci,Yuennan:2025kde,Oikonomou:2025xms,Oikonomou:2025htz,Odintsov:2025jky,Aoki:2025ywt,Gialamas:2025kef,Gialamas:2025ofz,Yuennan:2025tyx,Pallis:2025nrv,Pallis:2025gii}, 
with a comprehensive review presented in~\cite{Kallosh:2025ijd}. 

In this work, we have revisit the quantum-corrected Higgs inflation scenario within the framework of a non-minimally coupled scalar field model and analyzed its phenomenological implications in light of the most recent observations. Section~\ref{II} outlines the theoretical formulation of the model, while Section~\ref{III} presents the derivation of the slow-roll parameters and the analytical expressions for the key inflationary observables, namely the scalar spectral index $n_s$ and the tensor-to-scalar ratio $r$. The results are then confronted with the most recent observational constraints from the Atacama Cosmology Telescope (ACT) and DESI collaborations. Finally, our conclusions and main findings are summarized in Section~\ref{Con}.

\section{Quantum-corrected Higgs Infltion revisited}\label{II}
In this section, we closely follow the approach presented in Refs.~\cite{Barvinsky:2008ia,DeSimone:2008ei,Barvinsky:2009ii,Bezrukov:2010jz,Steinwachs:2011zs,Steinwachs:2013tr,Bezrukov:2012hx}, see also a review \cite{Martin:2013tda}. Let us begin with a general class of cosmological models characterized by a non-minimal coupling, whose action is given by
\begin{eqnarray}
S &=&\int \mathrm{d}^4x\,\sqrt{g}\Bigg[U(\varphi)R - \frac{G(\varphi)}{2}\,\nabla_{\mu}\Phi^{a}\nabla^{\mu}\Phi_{a} \nonumber\\&&\quad\quad\quad\quad\quad- V(\varphi)\Bigg].
\label{ONAction}
\end{eqnarray}
The scalar fields $\Phi^{a}$ ($a = 1, \ldots, N$) form a multiplet with a rigid internal $O(N)$ symmetry, and their internal indices are raised and lowered using the flat metric $\delta_{ab}$. Distinct models within this class are determined by specific forms of the functions $U(\varphi)$, $G(\varphi)$, and $V(\varphi)$, as well as by the number of scalar components $N$. To preserve the internal $O(N)$ invariance, these functions must depend solely on the modulus $
\varphi := \sqrt{\Phi^{a}\Phi_{a}}$, that is, gravity interacts with a multiplet of real scalar fields. The central idea of Higgs inflation is that the Standard Model (SM) Higgs boson itself plays the role of the inflaton field. To ensure agreement with observational data, a large non-minimal coupling to gravity of the form $\xi\varphi^2R$ is introduced, with $\xi \simeq 10^4$. The coeﬃcient functions $U(\varphi),\, V(\varphi)$ and $G(\varphi)$ together with their classical
parts contain one-loop radiative corrections of the form \cite{Barvinsky:2008ia}
\begin{eqnarray}
U(\varphi)&=& U_{\text{tree}}+ U_{1\text{-loop}}\nonumber\\&=& \frac{1}{2}(M_{p}^2 + \xi\varphi^2)+\frac{\varphi^2}{32\pi^2}\mathbf{C}\ln\frac{\varphi^2}{\mu^2},\label{U}\\
V(\varphi)&=&V_{\text{tree}}+V_{1\text{-loop}} \nonumber\\&=& \frac{\lambda}{4}(\varphi^2 - \nu^2)^2+\frac{\lambda\varphi^4}{128\pi^2}\mathbf{A}\ln\frac{\varphi^2}{\mu^2},\label{V}\\G(\varphi)&=&G_{\text{tree}}+G_{1\text{-loop}}\nonumber\\&=&1+\frac{1}{32\pi^2}\mathbf{E}\ln\frac{\varphi^2}{\mu^2},
\label{G}
\end{eqnarray}
The matter sector, representing the interaction part of the SM, can be schematically expressed as
\begin{eqnarray}
\mathcal{L}_{\text{int}}^{\text{SM}}
 &=& -\sum_{\chi}\frac{\lambda_{\chi}}{2}\chi^2\varphi^2
   - \sum_{A}\frac{g_{A}^2}{2}A_{\mu}^2\varphi^2
   \nonumber\\&&- \sum_{\Psi}y_{\Psi}\varphi\,\bar{\Psi}\Psi,
\end{eqnarray}
where $\lambda_{\chi}$, $g_{A}$, and $y_{\Psi}$ denote the scalar, gauge, and Yukawa couplings, respectively. The dominant mass contributions arise from heavy SM particles:
\begin{align}
m_{W^{\pm}}^2 = \frac{g^2}{4}\varphi^2,\,\,
m_Z^2 = \frac{(g^2 + g'^2)}{4}\varphi^2,\,\,
m_t^2 = \frac{y_t^2}{2}\varphi^2.
\label{Masses}
\end{align}
At high energies, the logarithmic coefficients $\mathbf{A}$, $\mathbf{C}$, and $\mathbf{E}$ receive contributions exclusively from Goldstone loops. Neglecting graviton loops (which are suppressed by the large effective Planck mass $M_{\text{P}}^{\text{eff}} = \sqrt{M_{\text{P}}^2 + \xi\varphi^2}$) and expanding in powers of $1/\xi$, we obtain
\begin{eqnarray}
\mathbf{A} &=& \frac{3}{8\lambda}\!\left[2g^4 + (g^2 + g'^2)^2 - 16y_t^4\right] \nonumber\\&&+ 6\lambda + \mathcal{O}(\xi^{-2}), \nonumber\\
\mathbf{C} &=& 3\xi\lambda + \mathcal{O}(\xi^0), \quad
\mathbf{E} = \mathcal{O}(\xi^{-2}).
\end{eqnarray}
The expression for $\mathbf{A}$ also includes the matter contributions in Eq.~(\ref{Masses}) to the effective potential. To connect with the standard slow-roll formalism, it is convenient to transform from the Jordan frame (JF) to the Einstein frame (EF), in which the action resembles that of general relativity minimally coupled to a scalar field (though the two are not physically equivalent). This transformation involves a conformal rescaling of the metric, a non-linear field redefinition, and a re-scaling of the potential \cite{Barvinsky:2009fy}
\begin{eqnarray}
g_{\mu\nu} &\rightarrow& {\bar g}_{\mu\nu}=\frac{2U}{M_{p}^2}g_{\mu\nu}, \\
\left(\frac{\mathrm{d}\chi}{\mathrm{d}\varphi}\right)^2
 &=& \frac{M_{p}^2}{2}\frac{G\,U + 3(U')^2}{U^2},\label{JFtoEF}
\end{eqnarray}
and
\begin{eqnarray}
V(\chi)= \Bigg(\frac{M_{p}^2}{2}\Bigg)^2\frac{V(\varphi)}{U^2(\varphi)}\Bigg|_{\varphi=\varphi(\chi)}.
\label{JFtoEF1}
\end{eqnarray}
We employ the effective action given by Eq.~(\ref{ONAction}) in the regime of large non-minimal coupling, $\xi \gg 1$, to investigate the inflationary dynamics at field values far exceeding those near the minimum of the classical potential, that is,
\[
\varphi^2 \gg \frac{M_P^2}{\xi} \gg \nu^2.
\]
Accordingly, we assume that the following two parameters are small:
\begin{align}
\frac{M_P^2}{\xi\,\varphi^2} \ll 1, \qquad
\frac{\mathbf{A}}{32\pi^2} \ll 1.
\label{smallness}
\end{align}
It is also natural to expect that other combinations of coupling constants have magnitudes comparable to $\mathbf{A}$, so that the second inequality in Eq.~(\ref{smallness}) likewise applies to them:
\[
\frac{1}{32\pi^2}(\mathbf{B},\,\mathbf{C},\,\mathbf{D},\,\mathbf{E},\,\mathbf{F}) \ll 1.
\]
The smallness conditions in Eq.~(\ref{smallness}) ensure the validity of the slow-roll approximation for the system described by Eqs.~(\ref{JFtoEF}-\ref{JFtoEF1}). Considering Eqs.(\ref{U}-\ref{G}) together with Eq.(\ref{JFtoEF}), we come up with
\begin{eqnarray}
\chi\simeq \sqrt{\frac{3}{2}} M_{p} \left(1+\frac{\bf C}{16 \pi ^2 \xi }\right) \log \left(\frac{\xi  \phi ^2}{M_{p}^2}\right)
\end{eqnarray}
Inverting the above expression gives
\begin{eqnarray}
\varphi\simeq \frac{e^{\frac{\chi }{\sqrt{6} M_{p}}}}{\sqrt{\xi }}-\frac{\left({\bf C}\, e^{\frac{\chi }{\sqrt{6} M_{p}}}\right)}{16 \left(\pi ^2 \xi ^{3/2}\right)}\frac{\chi }{\sqrt{6} M_{p}}\,.\label{var}
\end{eqnarray}
Consider the EF potential given in Eq.(\ref{JFtoEF1}) and in this frame, all cosmological observables can be expressed in terms of the EF effective potential \cite{Barvinsky:2009fy}
\begin{align}
V(\chi)
\simeq \frac{\lambda M_{\text{P}}^4}{4\xi^2}\left(1 - \frac{2M_{\text{P}}^2}{\xi\varphi^2}
+ \frac{\mathbf{A}_{\text{I}}}{16\pi^2}\ln\frac{\varphi}{\mu}\right),\label{JF}
\end{align}
where $\mathbf{A}_{\text{I}} = \mathbf{A} - 12\lambda$ denotes the inflationary anomalous scaling, incorporating the quantum Goldstone corrections contained in $\mathbf{C}$. Since we treat  $\bf A_I$ and $\lambda$ as free parameters, our setup is essentially the same as Ref.\cite{Gialamas:2025kef}, but parametrized in a different way. In fact, our result given in Eq.(\ref{JF}) can be easily mapped into Eq.(8) of Ref.\cite{Gialamas:2025kef}  when the
latter is expanded in the $\xi \gg 1$ limit at the order $\xi^{-3}$ and the
term $\delta\,\xi^{-3}$ is neglected. The above expression allows us to find the expression of the potential in the Einstein frame. Inserting Eq.(\ref{var}) into Eq.(\ref{JF}), one
obtains
\begin{align}
V(\chi)
\simeq \frac{\lambda M_{\text{P}}^4}{4\xi^2}\left(1-2 e^{-\frac{\sqrt{\frac{2}{3}} \chi }{M_{p}}}+\frac{{\bf A_{1}} \chi }{16 \sqrt{6} \pi^2 M_{p}}\right),\label{VE}
\end{align}
This formula coincides with Eq.(5.11) of Ref.\cite{Martin:2013tda}. Given Eq.(\ref{VE}), we can now proceed to the slow-roll analysis and the two first slow-roll
parameters can be written as
\begin{eqnarray}
\epsilon &\simeq& \frac{4}{3} e^{-\frac{2 \sqrt{\frac{2}{3}} \chi }{M_{p}}} \left(1+\frac{{\bf A_{I}} e^{\frac{\sqrt{\frac{2}{3}} \chi }{M_{p}}}}{64 \pi ^2}\right)^2\nonumber\\&\simeq& \frac{4}{3}\frac{M_{p}^{4}}{\xi^{2}\varphi^{4}}\Bigg(1+\frac{\varphi^{2}}{\varphi^{2}_{I}}\Bigg)^{2}\,,
\end{eqnarray}
where $\varphi^{2}_{I}\equiv 64\pi^{2}/{(\bf A_{I})}\xi$ and
\begin{eqnarray}
\eta\simeq -\frac{4}{3} e^{-\frac{2 \sqrt{\frac{2}{3}} \chi }{M_{p}}} \simeq -\frac{4}{3}\frac{M_{p}^{4}}{\xi^{2}\varphi^{4}}\,.
\end{eqnarray}
Following analysis presented in Ref.\cite{Barvinsky:1998rn}, the e-folding number of the inflation stage beginning with $\varphi$ and ending at $\varphi_{\rm end}$ can be determined to yield
\begin{eqnarray}
{\cal N} &=& -\int^{\varphi}_{\varphi_{\rm end}}{\rm d}\varphi'\frac{3H^{2}(\varphi')}{F(\varphi')}\nonumber\\&\simeq& \frac{48 \pi^2}{\bf A_I}\log \left(1+\frac{\varphi^{2}}{\varphi^{2}_{I}}\right)\,,
\end{eqnarray}
with a new function $F(\varphi)$ is defined as
\begin{eqnarray}
F=\frac{2VU'-V'U}{GU+3U^{'2}}\,.
\end{eqnarray}
Inverting the above formula gives
\begin{eqnarray}
\frac{\phi^2}{\phi_{0}^2}\simeq e^{x}-1\quad{\rm with}\quad x\equiv \frac{{\bf A_I} {\cal N}}{48 \pi^2}\,.
\end{eqnarray}
We can determine the Fourier power spectrum of the scalar metric perturbation from the relation
\begin{align}
\Delta^2 
= \frac{V}{24\pi^2 M_{p}^4 \varepsilon}\sim 2.1 \times 10^{-9},
\end{align}
where the right-hand side is evaluated at the time of first horizon crossing, $k = aH$. This condition connects the comoving wavenumber $k^{-1}$ with the number of e-folds $N$. Accordingly, the power spectrum takes the form
\begin{align}
\Delta^2(k) =
\frac{{\cal N}^2}{72\pi^2}\,
\frac{\lambda}{\xi^2}\,
\left(\frac{e^{x} - 1}{x\,e^{x}}\right)^2.
\label{zeta}
\end{align}
This expression is in agreement with Eq.(30) of Ref.\cite{Barvinsky:2008ia}. With the observed value $\Delta^2(k)\sim 2.1\times 10^{-9}$ \cite{Planck:2018jri}, a relation of $\tfrac{\lambda}{\xi^2}$ can be straightforwardly determined to obtain for ${\cal N}=60$:
\begin{eqnarray}
\frac{\lambda}{\xi^2}\sim 4.145\times 10^{-10} \Bigg(\frac{x e^{x}}{e^{x}-1}\Bigg)^{2}\,.
\end{eqnarray}
It should be noted that early renormalization-group analyses of the scalar-graviton coupling (e.g.Ref.\cite{Muta:1991mw}) indicate that $\xi$ remains moderate, close to the conformal value $1/6$. In the present inflationary framework, however, the large value $\xi \gg 1$ is not a result of RG running but follows from the CMB amplitude normalization, which fixes the combination $\lambda/\xi^{2} \sim 10^{-10}$. Hence, $\xi$ is large purely as a phenomenological requirement rather than a theoretical divergence of the coupling. The spectral index of the curvature perturbation is given by
\begin{align}
n_s \equiv 1 + \frac{d\ln \Delta^2}{d\ln k}
= 1 - 6\varepsilon(\chi) + 2\eta(\chi),
\end{align}
while the tensor-to-scalar ratio is expressed as $r = 16\varepsilon(\chi)$. These quantities can be explicitly written as
\begin{align}
n_s &= 1 - \frac{2}{{\cal N}}\,\frac{x}{e^{x} - 1}, \label{ns}\\
r &= \frac{12}{{\cal N}^2}\,
\left(\frac{x e^{x}}{e^{x} - 1}\right)^2. \label{r}
\end{align}
These approximate expressions match Eqs.(32)
and (34) of Ref.\cite{Barvinsky:2008ia}. It follows that in the limit $x \ll 1$, the spectral index simplifies to $n_s \simeq 1 - \frac{2}{{\cal N}}$,
which is identical to the prediction of the $m^{2}\varphi^{2}$ and $R + R^{2}/(6M^{2})$ inflationary models. Because the tensor-to-scalar ratio $r$ scales as ${\cal N}^{-2}$, its magnitude is considerably smaller than that obtained in the $m^{2}\varphi^{2}$ and $\lambda\varphi^{4}/4$ scenarios. Nevertheless, for $x \ll 1$, it exactly matches the result for the $f(R) = R + \tfrac{R^{2}}{6M^{2}}$ model~\cite{Starobinsky:1980te}.

\section{Confrontation with ACT data}\label{III}
Let us now compare the spectral index $n_s$ (\ref{ns}) to the present observational
data. Using the ACT+DESI (P-ACT-LB) constraints from \cite{ACT:2025fju,ACT:2025tim} at the $2\sigma$ confidence level, we get $0.567<x/(e^{x} - 1)<0.975$ or $0.0504<x<1.045$. Therefore, it follows that
\begin{eqnarray}
0.398<&{\bf A_{I}}&<8.248\,,\\
3.50\times 10^{-3}<&r&<8.65\times 10^{-3}\,.\label{rr}
\end{eqnarray}
We observe that the predicted values of $r$ in this work are in excellent agreement with the updated Planck constraint on the tensor-to-scalar ratio, $r<0.036$ at 95\% confidence level~\cite{BICEP:2021xfz}. The corresponding predictions for $r$ given in Eq.(\ref{rr}) can be tested by future precise cosmic microwave background (CMB) experiments. For instance, CORE \cite{COrE:2011bfs}, AliCPT \cite{Li:2017drr}, LiteBIRD \cite{Matsumura:2013aja}, and CMB-S4 \cite{Abazajian:2019eic} are
expected to achieve sensitivity to $r\sim {\cal O}(10^{-3})$. Using the above relations, a parameter  $\lambda/\xi^{2}$ can be constrained to yield
\begin{eqnarray}
4.36 \times 10^{-10}<\frac{\lambda}{\xi^2}<10.77\times 10^{-10}\,.
\end{eqnarray}
Considering the analyses in Refs.~\cite{Spokoiny:1984bd,Futamase:1987ua,Salopek:1988qh,Fakir:1990eg,Barvinsky:1994hx,Barvinsky:1998rn,Bezrukov:2007ep}, the predicted ratio $\tfrac{\lambda}{\xi^2}$ derived in this study is still consistent with previous findings, that is $\tfrac{\lambda}{\xi^2}\sim 10^{-10}$. 
\begin{figure}
\includegraphics[width=8 cm]{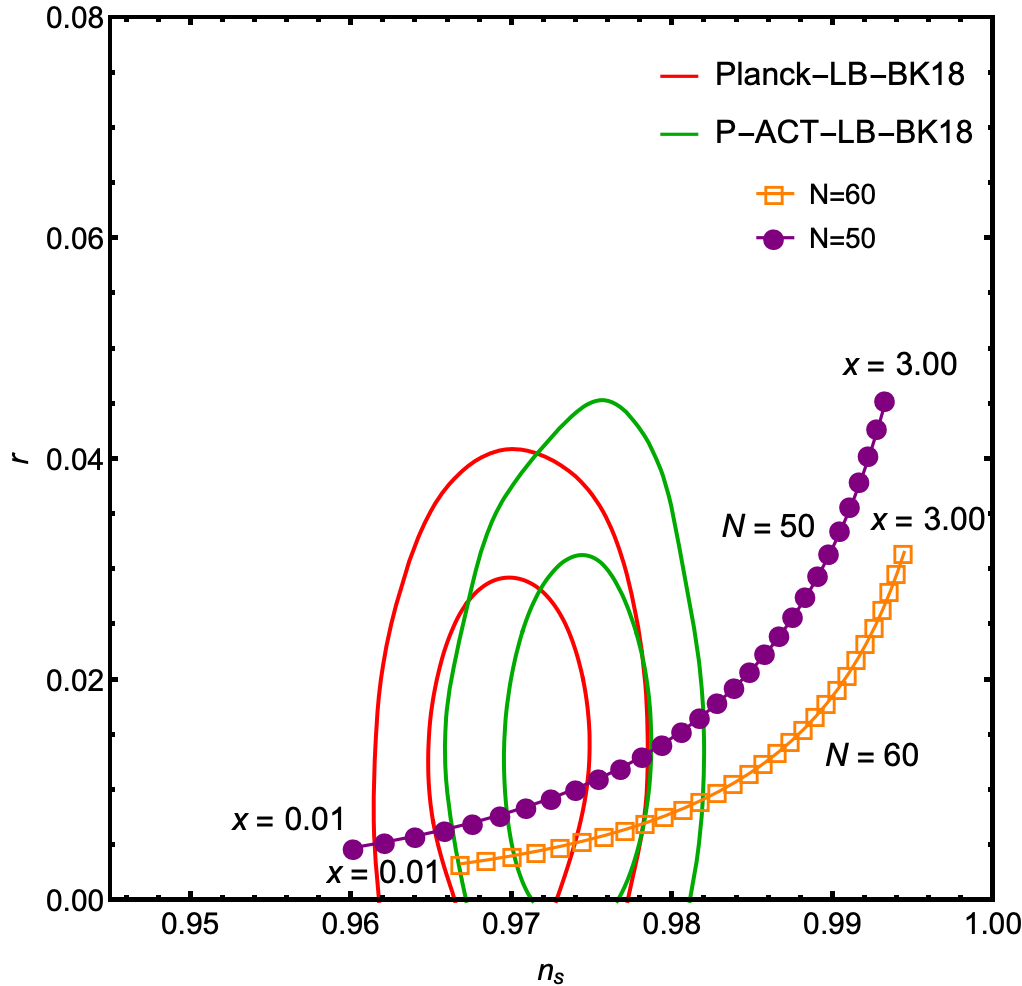}
\caption{Constraints on the scalar and tensor primordial power spectra, shown in the $r-n_{s}$
parameter space. The bounds on $r$ are primarily determined by the BK18 observations, whereas the limits on $n_s$ are set by Planck (red) and P-ACT (green) data. We fix $N=50,\,60$ and vary a parameter $x\equiv \tfrac{{\bf A} {\cal N}}{48 \pi^2}$.}\label{rns}
\end{figure}
From Fig.(\ref{rns}), we observe that Quantum-corrected Higgs Infltion scenario yield an improved agreement with the latest observational data, particularly when the inflaton field is allowed to couple non-minimally to gravity. In the limit $x\ll 1$, our framework reproduces the predictions of the Starobinsky $R^2$ model \cite{Starobinsky:1980te}, as well as those of Higgs and Higgs-like inflationary scenarios \cite{Kaiser:1994vs,Bezrukov:2007ep,Bezrukov:2008ej}. More specifically, for ${\cal N}=60,\,x={\bf A_{I}}{\cal N}/(48 \pi^{2})=0.500$, our model predicts $n_{s}=0.9743$ and $r=5.38\times 10^{-3}$, values that are in excellent agreement with the most recent observations  \cite{ACT:2025fju,ACT:2025tim}.

\section{Conclusions}\label{Con}
In this work, we have revisited the quantum-corrected Higgs inflation scenario within the framework of a non-minimally coupled scalar field model and analyzed its phenomenological implications in light of the most recent ACT and DESI (P-ACT-LB) observations. Starting from a one-loop effective action, we derived analytic expressions for the inflationary observables and demonstrated how the radiative corrections, encoded in the anomalous scaling parameter ${\bf A_I}$, modify the predictions for the spectral index $n_s$ and the tensor-to-scalar ratio $r$.

Our analysis shows that the quantum-corrected potential naturally yields inflationary predictions that remain well within the current observational bounds. For ${\cal N} = 60$, the model predicts $n_s \simeq 0.9743$ and $r \simeq 5.4 \times 10^{-3}$, in excellent agreement with the latest ACT+DESI (P-ACT-LB) data and fully consistent with the Planck 2018 limit $r < 0.036$. The inclusion of quantum corrections effectively shifts the predictions toward higher $n_s$ values, thereby improving compatibility with the ACT observations while keeping $r$ at values testable by near-future experiments. The derived constraint 
\[
4.36 \times 10^{-10} < \frac{\lambda}{\xi^2} < 10.77 \times 10^{-10},
\]
demonstrates that the quantum-corrected Higgs inflation framework remains robust and predictive. The results highlight that radiative effects play a crucial role in refining inflationary models based on the Standard Model Higgs sector. Upcoming high-precision CMB polarization missions such as CORE \cite{COrE:2011bfs}, AliCPT \cite{Li:2017drr}, LiteBIRD \cite{Matsumura:2013aja}, and CMB-S4 \cite{Abazajian:2019eic}, which aim for sensitivities of $r \sim 10^{-3}$, will be able to probe the parameter space outlined here, potentially providing decisive evidence in favor of or against quantum-corrected Higgs inflation.


\begin{thebibliography}{99}

\bibitem{ACT:2025fju}
T.~Louis \textit{et al.} [ACT],
[arXiv:2503.14452 [astro-ph.CO]].

\bibitem{ACT:2025tim}
E.~Calabrese \textit{et al.} [ACT],
[arXiv:2503.14454 [astro-ph.CO]].

\bibitem{DESI:2024uvr}
A.~G.~Adame \textit{et al.} [DESI],
JCAP \textbf{04} (2025), 012 

\bibitem{DESI:2024mwx}
A.~G.~Adame \textit{et al.} [DESI],
JCAP \textbf{02} (2025), 021

\bibitem{Planck:2018jri}
Y.~Akrami \textit{et al.} [Planck],
Astron. Astrophys. \textbf{641} (2020), A10

\bibitem{Starobinsky:1980te}
A.~A.~Starobinsky,
Phys. Lett. B \textbf{91} (1980), 99-102

\bibitem{Sato:1981qmu}
K.~Sato,
Mon. Not. Roy. Astron. Soc. \textbf{195} (1981) no.3, 467-479

\bibitem{Guth:1980zm}
A.~H.~Guth,
Phys. Rev. D \textbf{23} (1981), 347-356

\bibitem{Linde:1981mu}
A.~D.~Linde,
Phys. Lett. B \textbf{108} (1982), 389-393

\bibitem{Albrecht:1982wi}
A.~Albrecht and P.~J.~Steinhardt,
Phys. Rev. Lett. \textbf{48} (1982), 1220-1223

\bibitem{Kallosh:2013tua}
R.~Kallosh, A.~Linde and D.~Roest,
Phys. Rev. Lett. \textbf{112} (2014) no.1, 011303

\bibitem{Kallosh:2013hoa}
R.~Kallosh and A.~Linde,
JCAP \textbf{07} (2013), 002

\bibitem{Kallosh:2013maa}
R.~Kallosh and A.~Linde,
JCAP \textbf{10} (2013), 033

\bibitem{Kallosh:2013yoa}
R.~Kallosh, A.~Linde and D.~Roest,
JHEP \textbf{11} (2013), 198

\bibitem{Kallosh:2014rga}
R.~Kallosh, A.~Linde and D.~Roest,
JHEP \textbf{08} (2014), 052

\bibitem{Kallosh:2015lwa}
R.~Kallosh and A.~Linde,
Phys. Rev. D \textbf{91} (2015), 083528

\bibitem{Roest:2015qya}
D.~Roest and M.~Scalisi,
Phys. Rev. D \textbf{92} (2015), 043525

\bibitem{Linde:2016uec}
A.~Linde,
JCAP \textbf{02} (2017), 028

\bibitem{Terada:2016nqg}
T.~Terada,
Phys. Lett. B \textbf{760} (2016), 674-680

\bibitem{Ueno:2016dim}
Y.~Ueno and K.~Yamamoto,
Phys. Rev. D \textbf{93} (2016) no.8, 083524

\bibitem{Odintsov:2016vzz}
S.~D.~Odintsov and V.~K.~Oikonomou,
Phys. Rev. D \textbf{94} (2016) no.12, 124026

\bibitem{Akrami:2017cir}
Y.~Akrami, R.~Kallosh, A.~Linde and V.~Vardanyan,
JCAP \textbf{06} (2018), 041

\bibitem{Dimopoulos:2017zvq}
K.~Dimopoulos and C.~Owen,
JCAP \textbf{06} (2017), 027


\bibitem{Kaiser:1994vs}
D.~I.~Kaiser,
Phys. Rev. D \textbf{52} (1995), 4295-4306

\bibitem{Bezrukov:2008ej}
F.~L.~Bezrukov, A.~Magnin and M.~Shaposhnikov,
Phys. Lett. B \textbf{675} (2009), 88-92

\bibitem{Bezrukov:2007ep}
F.~L.~Bezrukov and M.~Shaposhnikov,
Phys. Lett. B \textbf{659} (2008), 703-706

\bibitem{Chang:2019ebx}
P.~W.~Chang, C.~W.~Chiang and K.~W.~Ng,
JHEP \textbf{04} (2020), 163


\bibitem{Karwan:2013iph}
K.~Karwan and P.~Channuie,
JCAP \textbf{06} (2014), 045

\bibitem{Channuie:2012bv}
P.~Channuie, J.~J.~Jorgensen and F.~Sannino,
Phys. Rev. D \textbf{86} (2012), 125035

\bibitem{Bezrukov:2011mv}
F.~Bezrukov, P.~Channuie, J.~J.~Joergensen and F.~Sannino,
Phys. Rev. D \textbf{86} (2012), 063513

\bibitem{Channuie:2011rq}
P.~Channuie, J.~J.~Joergensen and F.~Sannino,
JCAP \textbf{05} (2011), 007

\bibitem{Channuie:2014ysa}
P.~Channuie,
Nucl. Phys. B \textbf{892} (2015), 429-448

\bibitem{Samart:2022pza}
D.~Samart, C.~Pongkitivanichkul and P.~Channuie,
Eur. Phys. J. ST \textbf{231} (2022) no.7, 1325-1344

\bibitem{Gialamas:2025kef}
I.~D.~Gialamas, A.~Karam, A.~Racioppi and M.~Raidal,
[arXiv:2504.06002 [astro-ph.CO]].

\bibitem{Kallosh:2025rni}
R.~Kallosh, A.~Linde and D.~Roest,
[arXiv:2503.21030 [hep-th]].


\bibitem{Gao:2025onc}
Q.~Gao, Y.~Gong, Z.~Yi and F.~Zhang,
[arXiv:2504.15218 [astro-ph.CO]].



\bibitem{Liu:2025qca}
L.~Liu, Z.~Yi and Y.~Gong,
[arXiv:2505.02407 [astro-ph.CO]].

\bibitem{Yogesh:2025wak}
Yogesh, A.~Mohammadi, Q.~Wu and T.~Zhu,
[arXiv:2505.05363 [astro-ph.CO]].

\bibitem{Yi:2025dms}
Z.~Yi, X.~Wang, Q.~Gao and Y.~Gong,
[arXiv:2505.10268 [astro-ph.CO]].


\bibitem{Peng:2025bws}
Z.~Z.~Peng, Z.~C.~Chen and L.~Liu,
[arXiv:2505.12816 [astro-ph.CO]].

\bibitem{Yin:2025rrs}
W.~Yin,
[arXiv:2505.03004 [hep-ph]].

\bibitem{Byrnes:2025kit}
C.~T.~Byrnes, M.~Cort\^es and A.~R.~Liddle,
[arXiv:2505.09682 [astro-ph.CO]].

\bibitem{Wolf:2025ecy}
W.~J.~Wolf,
[arXiv:2506.12436 [astro-ph.CO]].

\bibitem{Aoki:2025wld}
S.~Aoki, H.~Otsuka and R.~Yanagita,
[arXiv:2504.01622 [hep-ph]].


\bibitem{Gao:2025viy}
Q.~Gao, Y.~Qian, Y.~Gong and Z.~Yi,
[arXiv:2506.18456 [gr-qc]].



\bibitem{Zahoor:2025nuq}
M.~Zahoor, S.~Khan and I.~A.~Bhat,
[arXiv:2507.18684 [astro-ph.CO]].


\bibitem{Ferreira:2025lrd}
E.~G.~M.~Ferreira, E.~McDonough, L.~Balkenhol, R.~Kallosh, L.~Knox
and A.~Linde,
[arXiv:2507.12459 [astro-ph.CO]].


\bibitem{Mohammadi:2025gbu}
A.~Mohammadi, Yogesh and A.~Wang,
[arXiv:2507.06544 [astro-ph.CO]].

\bibitem{Choudhury:2025vso}
S.~Choudhury, G.~Bauyrzhan, S.~K.~Singh and K.~Yerzhanov,
[arXiv:2506.15407 [astro-ph.CO]].

\bibitem{Odintsov:2025eiv}
S.~D.~Odintsov and V.~K.~Oikonomou,
Phys. Lett. B \textbf{870} (2025), 139907

\bibitem{Odintsov:2025wai}
S.~D.~Odintsov and V.~K.~Oikonomou,
Phys. Lett. B \textbf{868} (2025), 139779


\bibitem{Q:2025ycf}
R.~D.~A.~Q., J.~Chagoya and A.~A.~Roque,
[arXiv:2508.13273 [gr-qc]].


\bibitem{Zhu:2025twm}
Y.~Zhu, Q.~Gao, Y.~Gong and Z.~Yi,
[arXiv:2508.09707 [astro-ph.CO]].

\bibitem{Kouniatalis:2025orn}
G.~Kouniatalis and E.~N.~Saridakis,
[arXiv:2507.17721 [astro-ph.CO]].


\bibitem{Hai:2025wvs}
M.~Hai, A.~R.~Kamal, N.~F.~Shamma and M.~S.~J.~Shuvo,
[arXiv:2506.08083 [hep-th]].


\bibitem{Dioguardi:2025vci}
C.~Dioguardi, A.~J.~Iovino and A.~Racioppi,
Phys. Lett. B \textbf{868} (2025), 139664

\bibitem{Yuennan:2025kde}
J.~Yuennan, P.~Koad, F.~Atamurotov and P.~Channuie,
[arXiv:2508.17263 [astro-ph.CO]].

\bibitem{Oikonomou:2025xms}
V.~K.~Oikonomou,
[arXiv:2508.19196 [gr-qc]].

\bibitem{Oikonomou:2025htz}
V.~K.~Oikonomou,
[arXiv:2508.17363 [gr-qc]].

\bibitem{Odintsov:2025jky}
S.~D.~Odintsov and V.~K.~Oikonomou,
[arXiv:2508.17358 [gr-qc]].

\bibitem{Aoki:2025ywt}
S.~Aoki, H.~Otsuka and R.~Yanagita,
[arXiv:2509.06739 [hep-ph]].

\bibitem{Gialamas:2025ofz}
I.~D.~Gialamas, T.~Katsoulas and K.~Tamvakis,
JCAP \textbf{09} (2025), 060

\bibitem{Yuennan:2025tyx}
J.~Yuennan, F.~Atamurotov and P.~Channuie,
[arXiv:2509.23329 [gr-qc]].

\bibitem{Pallis:2025nrv}
C.~Pallis,
Phys. Lett. B \textbf{868} (2025), 139739

\bibitem{Pallis:2025gii}
C.~Pallis,
JCAP \textbf{09} (2025), 061

\bibitem{Kallosh:2025ijd}
R.~Kallosh and A.~Linde,
[arXiv:2505.13646 [hep-th]].

\bibitem{Barvinsky:2008ia}
A.~O.~Barvinsky, A.~Y.~Kamenshchik and A.~A.~Starobinsky,
JCAP \textbf{11} (2008), 021
\bibitem{DeSimone:2008ei}
A.~De Simone, M.~P.~Hertzberg and F.~Wilczek,
Phys. Lett. B \textbf{678} (2009), 1-8

\bibitem{Barvinsky:2009ii}
A.~O.~Barvinsky, A.~Y.~Kamenshchik, C.~Kiefer, A.~A.~Starobinsky and C.~F.~Steinwachs,
Eur. Phys. J. C \textbf{72} (2012), 2219

\bibitem{Bezrukov:2010jz}
F.~Bezrukov, A.~Magnin, M.~Shaposhnikov and S.~Sibiryakov,
JHEP \textbf{01} (2011), 016

\bibitem{Steinwachs:2011zs}
C.~F.~Steinwachs and A.~Y.~Kamenshchik,
Phys. Rev. D \textbf{84} (2011), 024026

\bibitem{Steinwachs:2013tr}
C.~F.~Steinwachs and A.~Y.~Kamenshchik,
AIP Conf. Proc. \textbf{1514} (2013) no.1, 161-164

\bibitem{Bezrukov:2012hx}
F.~Bezrukov, G.~K.~Karananas, J.~Rubio and M.~Shaposhnikov,
Phys. Rev. D \textbf{87} (2013) no.9, 096001

\bibitem{Martin:2013tda}
J.~Martin, C.~Ringeval and V.~Vennin,
Phys. Dark Univ. \textbf{5-6} (2014), 75-235

\bibitem{Barvinsky:2009fy}
A.~O.~Barvinsky, A.~Y.~Kamenshchik, C.~Kiefer, A.~A.~Starobinsky and C.~Steinwachs,
JCAP \textbf{12} (2009), 003

\bibitem{Barvinsky:1998rn}
A.~O.~Barvinsky and A.~Y.~Kamenshchik,
Nucl. Phys. B \textbf{532} (1998), 339-360

\bibitem{BICEP:2021xfz}
P.~A.~R.~Ade \textit{et al.} [BICEP and Keck],
Phys. Rev. Lett. \textbf{127} (2021) no.15, 151301

\bibitem{COrE:2011bfs}
F.~R.~Bouchet \textit{et al.} [COrE],
[arXiv:1102.2181 [astro-ph.CO]].

\bibitem{Li:2017drr}
H.~Li, S.~Y.~Li, Y.~Liu, Y.~P.~Li, Y.~Cai, M.~Li, G.~B.~Zhao, C.~Z.~Liu, Z.~W.~Li and H.~Xu, \textit{et al.}
Natl. Sci. Rev. \textbf{6} (2019) no.1, 145-154

\bibitem{Matsumura:2013aja}
T.~Matsumura, Y.~Akiba, J.~Borrill, Y.~Chinone, M.~Dobbs, H.~Fuke, A.~Ghribi, M.~Hasegawa, K.~Hattori and M.~Hattori, \textit{et al.}
J. Low Temp. Phys. \textbf{176} (2014), 733

\bibitem{Abazajian:2019eic}
K.~Abazajian, G.~Addison, P.~Adshead, Z.~Ahmed, S.~W.~Allen, D.~Alonso, M.~Alvarez, A.~Anderson, K.~S.~Arnold and C.~Baccigalupi, \textit{et al.}
[arXiv:1907.04473 [astro-ph.IM]].

\bibitem{Spokoiny:1984bd}
B.~L.~Spokoiny,
Phys. Lett. B \textbf{147} (1984), 39-43

\bibitem{Futamase:1987ua}
T.~Futamase and K.~i.~Maeda,
Phys. Rev. D \textbf{39} (1989), 399-404

\bibitem{Salopek:1988qh}
D.~S.~Salopek, J.~R.~Bond and J.~M.~Bardeen,
Phys. Rev. D \textbf{40} (1989), 1753

\bibitem{Fakir:1990eg}
R.~Fakir and W.~G.~Unruh,
Phys. Rev. D \textbf{41} (1990), 1783-1791

\bibitem{Barvinsky:1994hx}
A.~O.~Barvinsky and A.~Y.~Kamenshchik,
Phys. Lett. B \textbf{332} (1994), 270-276

\bibitem{Barvinsky:1998rn}
A.~O.~Barvinsky and A.~Y.~Kamenshchik,
Nucl. Phys. B \textbf{532} (1998), 339-360

\bibitem{Muta:1991mw}
T.~Muta and S.~D.~Odintsov,
Mod. Phys. Lett. A \textbf{6} (1991), 3641-3646

\end{thebibliography}
\end{document}